\pdfoutput=1
\documentclass[aps,twocolumn,superscriptaddress,letterpaper]{revtex4}
\usepackage{graphicx,amssymb,bm,natbib,amsfonts,amsmath}
\usepackage{hyperref}

\begin{document}

\title{Ferromagnetic Enhancement of CE-type Spin Ordering in (Pr,Ca)MnO$_3$}

\author{S.Y. Zhou}
\altaffiliation{Correspondence should be sent to szhou@lbl.gov, ychuang@lbl.gov and rwschoenlein@lbl.gov.}
\affiliation{Advanced Light Source, Lawrence Berkeley National Laboratory, Berkeley, CA 94720, USA}
\affiliation{Materials Sciences Division, Lawrence Berkeley National Laboratory, Berkeley, CA 94720, USA}
\author{Y. Zhu}
\affiliation{Materials Sciences Division, Lawrence Berkeley National Laboratory, Berkeley, CA 94720, USA}
\affiliation{Department of Applied Science, University of California, Davis, CA 95616,USA}
\author{M.C. Langner}
\affiliation{Materials Sciences Division, Lawrence Berkeley National Laboratory, Berkeley, CA 94720, USA}
\author{Y.-D. Chuang}
\altaffiliation{Correspondence should be sent to szhou@lbl.gov, ychuang@lbl.gov and rwschoenlein@lbl.gov.}
\affiliation{Advanced Light Source, Lawrence Berkeley National Laboratory, Berkeley, CA 94720, USA}
\author{P. Yu}
\affiliation{Department of Physics, University of California, Berkeley, CA 94720, USA}
\author{W.L. Yang}
\affiliation{Advanced Light Source, Lawrence Berkeley National Laboratory, Berkeley, CA 94720, USA}
\author{A.G. Cruz Gonzalez}
\affiliation{Advanced Light Source, Lawrence Berkeley National Laboratory, Berkeley, CA 94720, USA}
\author{N. Tahir}
\affiliation{Advanced Light Source, Lawrence Berkeley National Laboratory, Berkeley, CA 94720, USA}
\affiliation{National Center for Physics, Islamabad, Pakistan}
\author{M. Rini}
\affiliation{Materials Sciences Division, Lawrence Berkeley National Laboratory, Berkeley, CA 94720, USA}
\author{Y.-H. Chu}
\affiliation{Department of Materials Science and Engineering, National Chiao Tung University, Hsingchu 30010, Taiwan}
\author{R. Ramesh}
\affiliation{Materials Sciences Division, Lawrence Berkeley National Laboratory, Berkeley, CA 94720, USA}
\affiliation{Department of Physics, University of California, Berkeley, CA 94720, USA}
\author{D.-H. Lee}
\affiliation{Materials Sciences Division, Lawrence Berkeley National Laboratory, Berkeley, CA 94720, USA}
\affiliation{Department of Physics, University of California, Berkeley, CA 94720, USA}
\author{Y. Tomioka}
\affiliation{Nanoelectronics Research Institute, National Institute of Advanced Industrial Science and Technology (AIST)
Tsukuba Central 4, 1-1-1 Higashi Tsukuba 305-8562, Japan}
\author{Y. Tokura}
\affiliation{Department of Applied Physics, University of Tokyo, Bunkyo-ku, Tokyo 113-8656, Japan}
\affiliation{Cross-Correlated Materials Research Group (CMRG) and Correlated Electron Research Group (CERG),
Advanced Science Institute, RIKEN, Wako 351-0198, Japan}
\author{Z. Hussain}
\affiliation{Advanced Light Source, Lawrence Berkeley National Laboratory, Berkeley, CA 94720, USA}
\author{R.W. Schoenlein}
\altaffiliation{Correspondence should be sent to szhou@lbl.gov, ychuang@lbl.gov and rwschoenlein@lbl.gov.}
\affiliation{Advanced Light Source, Lawrence Berkeley National Laboratory, Berkeley, CA 94720, USA}
\affiliation{Materials Sciences Division, Lawrence Berkeley National Laboratory, Berkeley, CA 94720, USA}

\date{\today}

\begin{abstract}
We present resonant soft X-ray scattering (RSXS) results from small band width manganites (Pr,Ca)MnO$_3$, which show that the CE-type spin ordering (SO) at the phase boundary is stabilized only below the canted antiferromagnetic transition temperature and enhanced by ferromagnetism in the macroscopically insulating state (FM-I). Our results reveal the fragility of the CE-type ordering that underpins the colossal magnetoresistance (CMR) effect in this system, as well as an unexpected {\it cooperative} interplay between FM-I and CE-type SO which is in contrast to the {\it competitive} interplay between the ferromagnetic metallic (FM-M) state and CE-type ordering.
\end{abstract}

\maketitle

The role of localized orderings and their interplay with competing ground states is central to understanding the novel physics of transition metal oxides which exhibit a variety of metal-insulator transitions, e.g.~high temperature superconductivity, colossal magnetoresistance (CMR) etc \cite{RMP98, TokuraSci}. In high temperature superconductors, the anomalous suppression of superconductivity around 1/8 doping in La$_{2-x}$Ba$_x$CuO$_4$ is attributed to the formation of static charge stripe ordering \cite{Tranquada}. In manganites \cite{DagottoRP, TokuraRPP06}, the competitive nature between ferromagnetic metallic (FM-M) and charge/orbital/spin ordered (CO/OO/SO) insulating states is best illustrated by the phase diagram shown in Fig.1(a), in which transition between different low temperature ground states can be driven by tuning the average ionic radius across the critical value of 1.33$\AA$ \cite{TokuraPRBtolerance}. Ferromagnetism in the FM-M state, through the double exchange interaction, tends to delocalize electrons, and is therefore in competition with electronic orderings. However, it is unclear whether the ferromagnetism mediated by the super-exchange interaction is also playing a competing role.  Elucidating their interplay is critical for understanding the low temperature CMR effect which involves the transition from a long range ordered insulating state to a FM-M state \cite{Tokura2CMR, DagottoNJP} in systems like (Pr,Ca)MnO$_3$ \cite{PCMOFernandez}, where the CMR effect is largest \cite{TomiokaPRB96} and the underlying physics is not well understood. In this letter, we present experimental results supporting the cooperative interplay between FM-I and spin ordering (SO) in 30$\%$ doped (Pr,Ca)MnO$_3$.

\begin{figure}
\includegraphics[width=8.8 cm] {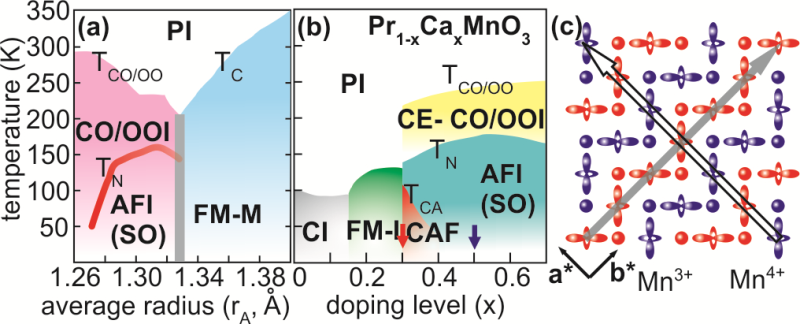}
\label{Figure 1}
\caption{(a, b) Phase diagrams as a function of (a) A-site average ionic radius and (b) doping. The CO/OOI, PI, CI, AFI and CAF denote charge/orbital ordered insulating, paramagnetic insulating, spin-canted insulating, antiferromagnetic insulating and canted antiferromagnetic states respectively. (c) Schematic drawing of the CE-type CO/OO/SO \cite{footnoteCE} at 50$\%$ doping. The arrows mark the directions for the OO and SO wave vectors. Red and blue colors represent opposite spins on the Mn sites.}
\end{figure}

\begin{figure*}
\includegraphics[width=13 cm] {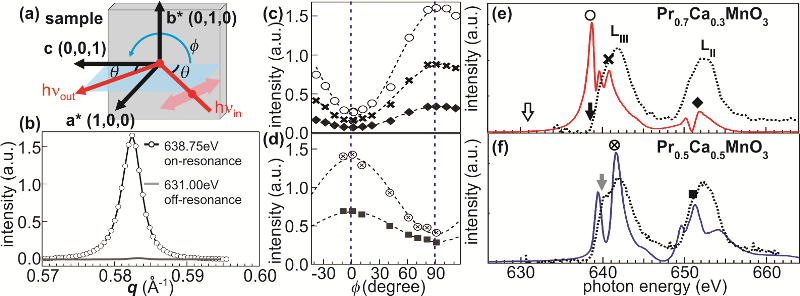}
\label{Figure 2}
\caption{(a) Schematic drawing of the experimental geometry. The azimuthal angle $\phi$ is defined to be 0 when c-axis is in the scattering plane. (b) Intensity of the ordering peak as a function of momentum transfer q measured on 30$\%$ doped (Pr, Ca)MnO$_3$ at 65 K with incident photon energies of 638.75 eV and 631 eV (block arrows in panel (e)) at $\phi$=90$^\circ$. (c,d) Azimuthal angular dependence of the ordering peak intensity on (c) 30$\%$ and (d) 50$\%$ doped samples at photon energies marked by symbols in panels (e,f). The dotted lines are fits to the data using a sin$^2(\phi-\phi_0)$ function. (e,f) Resonance profiles measured at 65 K on (e) 30$\%$ doped sample with $\phi$= 90$^\circ$ and (f) 50$\%$ doped sample with $\phi$=0$^\circ$ respectively. The dotted lines are the XAS spectra measured in the total fluorescence yield (TFY) mode (rescaled by 100 times).}
\end{figure*}

Resonant soft X-ray scattering (RSXS) experiments were carried out at Beamline 8 of the Advanced Light Source (ALS) at Lawrence Berkeley National Laboratory using a two-circle diffractometer in a horizontal scattering geometry (see Fig.2(a)). Single crystal (Pr,Ca)MnO$_3$ samples with 30$\%$ (Pr$_{0.7}$Ca$_{0.3}$MnO$_3$) and 50$\%$ (Pr$_{0.5}$Ca$_{0.5}$MnO$_3$) doping grown by traveling floating zone method were cut and polished to have optically flat surfaces with surface normal along the (1,0,0) direction in the orthorhombic setting \cite{footnote0}. Because of the twin domains in the crystal, the orbital ordering (OO) and SO peaks at (0,1/2,0) and (1/2,0,0) can both be observed at similar scattering angles. However, their contributions can be distinguished via azimuthal angular dependence, temperature dependence and the slightly different ordering wave vectors \cite{footnote0}.

Resonant X-ray scattering is a direct probe for electronic ordering \cite{MurakamiLSMOPRL98}. In particular RSXS at the Mn-L edge not only provides elemental sensitivity but also enhances the weak electronic contributions from the Mn 3d orbitals \cite{CastletonPRB00, WilkinsPRL03, DhesiPRL04, ThomasPRL}. The resonance effect can be seen in Fig.2(b) where the diffraction peak from CE-type ordering is greatly enhanced when the incident photon energy is tuned on resonance. The observed ordering peak is very sharp - the full width half maximum $\Delta$q is smaller than 0.003\AA$^{-1}$, from which the correlation length $\xi$=2$\pi$/$\Delta$q is estimated to be larger than 2000\AA. This is surprising because the spectrum was taken from a 30$\%$ doped sample in the canted antiferromagnetic phase and it shows the ordering still has long range correlation even in close proximity to the FM-I state. 

To understand the RSXS results from the 30$\%$ doped sample, we compare them with those from a 50$\%$ doped sample where the ordering is most prominent. Both samples were measured under identical experimental conditions and yet the RSXS data show striking differences. First of all, the maximum intensity occurs at different azimuthal angles - $\phi$=0$^\circ$ (c-axis in the scattering plane) for 50$\%$ doped sample and $\phi$=90$^\circ$ (c-axis perpendicular to the scattering plane) for 30$\%$ doped sample (see Fig.2(c,d)), and the difference does not depend on the incident photon energies. Second, striking differences can also be seen in the resonance profiles - intensity of the diffraction peak as a function of incident photon energy, even though the X-ray absorption spectra (XAS, see black dotted curves in Figs.2(e,f)) are overall very similar, aside from a slightly more pronounced shoulder at $\approx$ 640eV for the 50$\%$ doped sample (indicated by the arrow in Fig.2(f)). Such small differences in XAS spectra are likely caused by the variation in Mn valency as more Mn$^{4+}$ states (fewer e$_g$ electrons) are expected in the 50$\%$ doped sample, and the resulting change in the local environment (multiplet) might favor the OO and SO differently, leading to different resonance profiles (Figs.2(e,f)). 
The differences in the azimuthal angular dependence and resonance profiles suggest that the electronic ordering in the 30$\%$ doped sample is very different from that in the 50$\%$ sample, albeit they share similar wave vectors.

\begin{figure*}
\includegraphics[width=12.4cm] {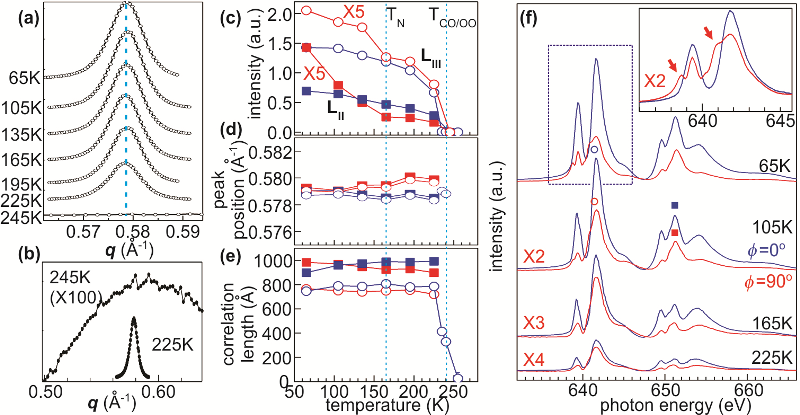}
\label{Figure 3}
\caption{Temperature dependence of (a,b) the diffraction peak at 641.6 eV with $\phi$=0$^\circ$, (c) extracted intensity (d) peak position (e) correlation length, and (f) resonance profiles on 50$\%$ doped sample. The blue (red) symbols and curves in panels (c-e) were taken with $\phi$=0$^\circ$ ($\phi$=90$^\circ$) respectively. The inset in panel (f) shows the zoom-in of the resonance profiles at L$_{III}$ edge. The circular and square symbols in panel (f) mark the photon energies at which data in panels (c-e) were taken.}
\end{figure*}

Although theoretical calculations of resonance profiles \cite{CastletonPRB00, DhesiPRL04, ThomasPRL} will be useful to provide insights on the different orderings in these two samples, the treatment of quantum interference between multiple valency states in RSXS processes has been a challenging model-dependent issue.  Thus to draw general (yet unambiguous) conclusions, we first performed a detailed study on the 50$\%$ doped sample (Fig.3). From 245K to 225K, the correlation length of the ordering increases quickly to $\approx$1000$\AA$ (Fig.3(e)). Below 225K, only the peak intensity (Fig.3(c)) shows strong temperature dependence while the ordering wave vector (Fig.3(d)) and correlation length (Fig.3(e)) remain nearly temperature-independent. This agrees with the general consensus that the ordering in the 50$\%$ doped sample is truly long range and energetically favorable.  We further conclude that the $\phi$=0$^\circ$ geometry is mainly sensitive to OO while the $\phi$=90$^\circ$ geometry has additional sensitivity to SO, based on the following observations. First, resonance profiles taken with $\phi$=0$^\circ$ (blue curves in Fig.3(f)) agree well with those from OO in (La,Sr)$_3$Mn$_2$O$_7$ which has only OO. Second, at $\phi$=90$^\circ$, below T$_N$ $\approx$ 165K there are two additional peaks (indicated by arrows in the inset of Fig.3(f)) in the resonance profiles at similar energies as those identified as SO peaks \cite{StaubPRB09}. The abrupt increase in the diffraction peak intensity at the L$_{II}$ edge (red square symbols in Fig.3(c)) as the temperature is lowered through T$_N$, also signals an additional SO contribution in the $\phi$=90$^\circ$ geometry. Finally, this assignment of the OO (SO) sensitivity also agrees with previous studies \cite{ThomasPRL, StaubPRB09}. 

\begin{figure*}
\includegraphics[width=13.2 cm] {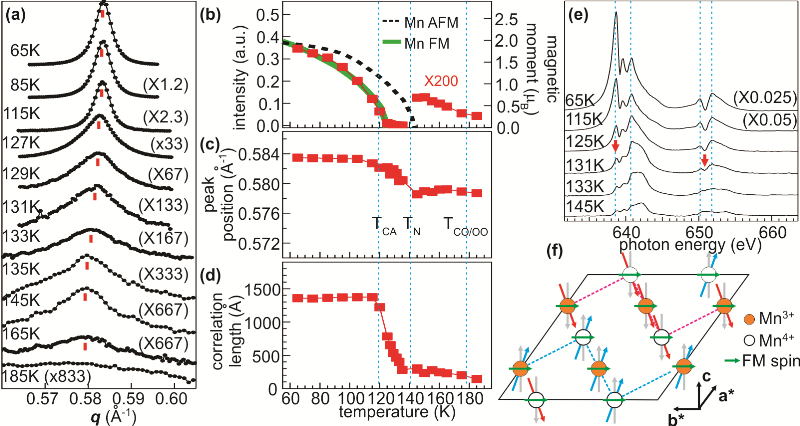}
\label{Figure 4}
\caption{Temperature dependence of (a) the diffraction peak (b) extracted intensity (c) peak position (d) correlation length and (e) resonance profiles measured on 30$\%$ doped sample at 652 eV with $\phi$=90$^\circ$. The green curve and the black broken line in panel (c) show the Mn Ferromagnetic moment and Mn antiferromagnetic moment (rescaled by 1.2) from neutron scattering measurements \cite{CoxPRB98}. (f) A schematic drawing of the spin orientation in the canted antiferromagnetic state.}
\end{figure*}

Having established the sensitivity to SO (OO) in the $\phi$=90$^\circ$ ($\phi$=0$^\circ$) geometry, the azimuthal angular dependence data shown in Fig.2(c,d) suggests that at 65 K, the 30$\%$ doped sample (with maximum intensity at $\phi$=90$^\circ$) is dominated by SO, while the 50$\%$ doped sample (with maximum intensity at $\phi$=0$^\circ$) is dominated by OO. Armed with such knowledge, we proceed to analyze the temperature dependence data from the 30$\%$ doped sample (Fig.4). Below T$_{CO/OO}$, a weak and broad OO peak is observed. Below T$_N$, not only does the peak intensity gradually increase, but also the peak position moves quickly to slightly larger wave vector and the correlation length increases. The change of peak position suggests that the SO, whose ordering vector is slightly larger by 0.005$\AA ^{-1}$, is gradually becoming dominant. Below T$_{CA}$$\approx$ 120K, the nearly temperature-independent peak position and correlation length, similar to what is observed in the 50$\%$ doped sample, signals the presence of long range SO. Significant changes are also observed in the resonance profiles between T$_N$ and T$_{CA}$ (Fig.4(e)). At 145K (above T$_N$) the resonance profile bears certain similarity to those of OO (cf. blue curves in Fig.3(e)). Around 130K, gradual development of a spectral dip at the L$_{II}$ edge and a low energy peak at $\approx$ 639eV is in agreement with the notion that SO is becoming dominant. However, unlike previous RSXS studies on 40$\%$ doped (Pr,Ca)MnO$_3$, increase of the SO peak in the 30$\%$ sample occurs mainly below T$_{CA}$ - 20K lower than the expected temperature T$_{N}$. This 20K difference points to the fragility of SO at this doping in proximity to FM-I phase and suggests that there is an additional mechanism that helps stabilizing it at lower temperature. 

The fragility of the CE-type ordering state can be understood by the charge disproportionation. In 30$\%$ doped sample where there are more Mn$^{3+}$ than Mn$^{4+}$, the additional e$_g$ electrons act as {\it disorder} for the CE-type OO, making it much shorter range. The antiferromagnetic SO will also be affected but to a much less extent, since not only the e$_g$ electron in Mn$^{3+}$ but also the three t$_{2g}$ electrons which are present in both Mn$^{3+}$ and Mn$^{4+}$ are involved in SO. This explains why in both 40$\%$ and 30$\%$ doped samples the SO peak is much stronger (by at least 100 time) than the OO peak. In addition, 30$\%$ doped sample also lies at the phase boundary between FM-I and CE-type SO (cf. phase diagram in Fig.1(b)). Right below T$_N$, the competition between FM-I with in-plane spins (below 30$\%$ doping) and SO with out-of-plane spins (above 30$\%$ doping) results in only fragile SO with short correlation length. The spins in the canted antiferromagnetic phase (red and blue arrows in Fig.4(f)) \cite{JirakJMMM85, TokuraPRB95, CoxPRB98}, to first order, can be viewed as a combination of FM-I with in-plane spin component (green arrows) and CE-type antiferromagnetic SO with out-of-plane spin component (gray arrows), either coexisting in the same region or phase separated in real space \cite{JirakJMMM85}. Although our results cannot distinguish either scenario conclusively, the spin canting seems to be a more straightforward picture. Such canting favors both FM-I and SO, and the cooperative interplay of ferromagnetism and CE-type ordering becomes possible.

The cooperative interplay between FM-I and CE-type SO in the canted antiferromagnetic phase is further supported by the strikingly similar temperature dependence between the SO peak intensity measured in RSXS and the ferromagnetic moment measured on the Mn sites from neutron scattering \cite{CoxPRB98} (Fig.4(b)). The cooperative or balanced interplay of ferromagnetism and CE-type ordering at T$_{CA}$ is analogous, for instance, to the formation of the layered (A-type) super-exchange mediated antiferromagnetic and metallic state for overdoped manganites, e.g., RE$_{1-x}$Sr$_x$MnO$_3$ (x = 0.55) (RE = Pr, Nd) \cite{TomiokaPRB03}, and is generic to doped manganites exhibiting rich phases upon doping. The {\it cooperative} interplay between the CE-type SO and FM-I state suggests that the mechanism behind the FM-I state in 30$\%$ doped (Pr,Ca)MnO$_3$ is likely to be mediated by super-exchange \cite{EndohPRL1999}, since double exchange tends to delocalize electrons and enhances the {\it competition} between ferromagnetism and ordering. 

In summary, our RSXS results show that although the CE-type OO is a robust ground state for 50$\%$ doped (Pr, Ca)MnO$_3$, the SO/OO is fragile in 30$\%$ doped sample due to the large charge disproportionation and the competition of FM-I and SO, and the SO is enhanced only by the development of Mn ferromagnetic moment in the canted antiferromagnetic phase. We believe that such delicate balance between competition (right below T$_N$) and cooperation (below T$_{CA}$) between FM-I and CE-type SO is directly relevant to the large CMR effect reported in this material.

\begin{acknowledgments}
The Advanced Light Source is supported by the Director, Office of Science, Office of Basic Energy Sciences, of the U.S. Department of Energy under Contract No. DE-AC02-05CH11231. This work was supported by the Director, Office of Science, Office of Basic Energy Sciences, the Materials Sciences and Engineering Division under the Department of Energy Contract No. DE-AC02-05CH11231. Y.H.C. is supported by NSC 099-2811-M-009-003.
\end{acknowledgments}


\begin {thebibliography} {99}

\bibitem{RMP98} Imada, M., et al., Rev. Mod. Phys. {\bf 70}, 1039 (1998).

\bibitem{TokuraSci} Tokura, Y. and Nagaosa, N., Science {\bf 288}, 462 (2000).

\bibitem{Tranquada} Tranquada, J.M. et al., Nature {\bf 375}, 561 (1995).

\bibitem{DagottoRP} Dagotto, E., et al, Phys. Rep. {\bf 344}, 1 (2001).

\bibitem{TokuraRPP06} Tokura, Y., Rep. Prog. Phys. {\bf 69}, 797 (2006).

\bibitem{TokuraPRBtolerance} Tomioka Y. and Tokura, Y., Phys. Rev. B {\bf 66}, 104416 (2002).

\bibitem{DagottoNJP} Dagotto, E., New J. Phys. {\bf 7}, 67 (2005).

\bibitem{Tokura2CMR} Tokura, Y., et al.,  Phys. Rev. Lett. {\bf 76}, 3184 (1996).

\bibitem{PCMOFernandez} Fernandez-Baca, J.A. et al., Phys. Rev. B {\bf 66}, 054434 (2002).

\bibitem{TomiokaPRB96} Tomioka, Y. et al., Phys. Rev. B {\bf 53}, R1689 (1996).

\bibitem{footnoteCE} We note that ``CE'' refers to the specific SO and here we use ``CE-type'' in a more general way to refer also to the CO and OO shown in Fig.1(c). We also note that there is a debate about the Mn valency (exact value of $\delta$ in Mn$^{3+\delta}$ and Mn$^{4-\delta}$).  For notation convenience, we will use Mn$^{3+}$ and Mn$^{4+}$ throughout this paper.










\bibitem{footnote0} For 30$\%$ doped sample, a=5.42$\AA$, b=5.47$\AA$ below 150K \cite{Radaelli1997, CoxPRB98}), and the SO wave vector is larger than the OO wave vector by $\approx$ 0.005 $\AA^{-1}$. 

\bibitem{MurakamiLSMOPRL98} Murakami, Y. et al., Phys. Rev. Lett. {\bf 80}, 1932 (1998).


\bibitem{CastletonPRB00} Castleton,C.W.M. and Altarelli, M., Phys. Rev. B {\bf 62}, 1033 (2000).

\bibitem{WilkinsPRL03} Wilkins,S.B. et al., Phys. Rev. Lett. {\bf 91}, 167205 (2003).

\bibitem{DhesiPRL04} Dhesi, S.S. et al., Phys. Rev. Lett. {\bf 92}, 056403 (2004).

\bibitem{ThomasPRL} Thomas, K.J. et al., Phys. Rev. Lett. {\bf 92}, 237204 (2004).



\bibitem{StaubPRB09} Staub, U. et al., Phys. Rev. B {\bf 79}, 224419 (2009).

\bibitem{CoxPRB98} Cox, D.E. et al. Phys. Rev. B {\bf 57}, 3305 (1998).



\bibitem{JirakJMMM85} Jirak,Z. et al., J. Mag. Mag. Mat. {\bf 53}, 153 (1985).

\bibitem{TokuraPRB95} Yoshizawa, H., et al. Phys. Rev. B {\bf 52}, R13145 (1995).

\bibitem{TomiokaPRB03} Tomioka, T. et cal. Phys. Rev. B {\bf 68}, 094417 (2003).

\bibitem{EndohPRL1999} Endoh, Y. et al. Phys. Rev. Lett. {\bf 82}, 4328 (1999).

\bibitem{Radaelli1997} Radaelli, P.G. et al., Phys. Rev. B {\bf 56}, 8265 (1997).








\end {thebibliography}
\end{document}